\documentclass[aps,pra,showpacs,twocolumn,superscriptaddress,amsmath,amssymb]{revtex4}

\usepackage{epsfig,afterpage}
\usepackage{graphicx}
\usepackage{amsfonts}
\usepackage{amssymb}
\usepackage{indentfirst}
\usepackage{amsmath,amsthm}
\usepackage{dsfont}

\usepackage{epsfig}
\usepackage{subfigure}

\usepackage{multirow}
\usepackage{pstricks}
\usepackage{lscape}

\usepackage{wasysym} 

\def\beq{\begin{equation}}
\def\eeq{\end{equation}}
\def\bea{\begin{eqnarray}}
\def\eea{\end{eqnarray}}
\def\bq{\begin{quote}}
\def\eq{\end{quote}}
\def\ben{\begin{enumerate}}
\def\een{\end{enumerate}}
\def\bit{\begin{itemize}}
\def\eit{\end{itemize}}
\def\nn{\nonumber}

\def\PEPS{PEPS}
\def\SGS{SGS}
\def\BSGS{B-SGS}

\begin{document}

\title{Sequentially generated states for the study of two dimensional systems}

\author{M. C. Ba\~nuls}\email{banulsm@mpq.mpg.de} 
\affiliation{Max-Planck-Institut f\"ur Quantenoptik,
Hans-Kopfermann-Str. 1, 85748 Garching, Germany.}
\author{D. P\'erez-Garc\'{\i}a}
\affiliation{Depto. An\'alisis Matem\'atico, Universidad Complutense de Madrid, 28040 Madrid, Spain.}
\author{M. M. Wolf}
\affiliation{Max-Planck-Institut f\"ur Quantenoptik,
Hans-Kopfermann-Str. 1, 85748 Garching, Germany.}
\author{F. Verstraete}
\affiliation{Fakult\"at f\"ur Physik, Universit\"at Wien, Boltzmanngasse 5, A-1090 Wien, Austria.}
\author{J.I. Cirac}
\affiliation{Max-Planck-Institut f\"ur Quantenoptik,
Hans-Kopfermann-Str. 1, 85748 Garching, Germany.}
\date{\today}

\begin{abstract}
Matrix Product States can be defined as the family of quantum
states that can be sequentially generated in a one--dimensional
system \cite{schon05seq}. We introduce a new family of states
which extends this definition to two dimensions. Like in Matrix
Product States, expectation values of few body observables can be
efficiently evaluated and, for the case of translationally
invariant systems, the correlation functions decay exponentially
with the distance. We show that such states are a subclass of
Projected Entangled Pair States and investigate their suitability
for approximating the ground states of local Hamiltonians.
\end{abstract}

\pacs{03.67.Mn, 02.70.-c, 75.10.Jm}

\maketitle

\section{Introduction}

The description of quantum many-body systems is a complex problem
due to the exponential growth of the dimension of the Hilbert
space with the number of particles. In many cases of physical
interest, however, states can be approximately described with a
small number of parameters. This is the reason for the success of
techniques such as the Density Matrix Renormalization Group
(DMRG)~\cite{white92dmrg}, or those based on Matrix Product State
(MPS) representations~\cite{perez07mps}. Those techniques take
advantage of the local character of physical interactions, which
favors states with a small amount of entanglement~\cite{vidal03eff}.
Their applicability is, however, limited to one-dimensional
systems.

MPS have a natural generalization to higher dimensional systems,
namely the Projected Entangled Pair States
(\PEPS)~\cite{verstraete04peps}. Both representations are
complete, i.e.\ any state of the Hilbert space can be written as a
MPS or PEPS, and they have an efficient description in terms of
the required number of parameters. However, they have very
different properties. For example, MPS can be efficiently
created~\cite{schon05seq} and classically
simulated~\cite{vidal03eff}, what makes them extremely useful for
the study of quantum 1D systems. However, creating and simulating
~\PEPS ~has been shown to be much harder~\cite{schuch07complex}.
Already computing local expectation values on PEPS has an
exponential cost in the general case. Nevertheless, they have
proved successful for studying the ground states properties of 2D
systems by means of an approximate
method~\cite{verstraete04peps,murg07hard}. In spite of having
polynomial cost in all the parameters involved, the consumption of
computational resources limits the application of those methods to
relatively small 2D systems \cite{murg07hard} or large ones but
with moderate precision ~\cite{jordan07infinite}.

One of the goals in the research with many--body systems is to
find other families of states providing an efficient description
of systems in two or higher dimensions, while keeping a more benign
behavior with regard to the determination of expectation values.
Those studies may find immediate applications in the numerical
studies of the physics of strongly correlated quantum systems.
Thus, in the last years, other classes of states and corresponding
variational methods have also been proposed to describe higher
dimensional
systems~\cite{vidal07mera,anders06graph,moukouri04heis,schuch07string,dawson07simu,hastings07belief}.

In this paper we present and discuss a new generalization of MPS
to two dimensions. The family of states introduced here is a
subfamily of \PEPS, specialized for 2D lattices: (i) which can be 
efficiently constructed and (ii) for which the expectation values
can be efficiently determined. These properties (inherited from
MPS) make such states candidates for variational algorithms that
search for ground states of local Hamiltonians.

The rest of the paper is organized as follows. In the next section we
review the definitions and properties of MPS and their PEPS generalization.
In section~\ref{sec:upeps} we introduce the new family of states by
presenting two alternative ways of extending
the MPS construction to 2D systems. In subsection~\ref{subsec:upeps_prop}
the main properties of such generalizations are discussed, whereas~\ref{subsec:upeps_algo}
shows their performance as ansatz for a variational algorithm.
We conclude with a discussion of these results in section~\ref{sec:conclu}.
The complete proof of the exponential decay of correlations in a translationally
invariant state is deferred to the appendix.

\section{Efficient representations of quantum many-body states}
\label{sec:effrep}

As discussed above, having a representation of quantum states 
that captures 
the essential entanglement features
turns out to be most desirable for the study of quantum many-body systems.
A good representation should additionally satisfy some
other properties.
It is not only necessary that the state can be 
described or well approximated in this manner, 
but also to be able to find such a description
and to determine physical quantities
in an efficient way.

Matrix Product States satisfy all these requirements for
one dimensional systems.
In the case of higher dimensions, the PEPS family
provides also an efficient representation of states, which 
by contrast results in a costly calculation of
physical quantities.
The following paragraphs review both families and their
properties in some detail.

\subsection{One dimensional systems: Matrix Product States}
\label{subsec:mps}

As already discussed, MPS constitute 
the paradigm of an efficient representation
for one dimensional quantum many-body systems~\cite{perez07mps}.
Here we recall the various ways in which they can be defined, as
well as their most significant properties.

\subsubsection{Definition}
\label{subsec:mps_def}

\begin{figure}
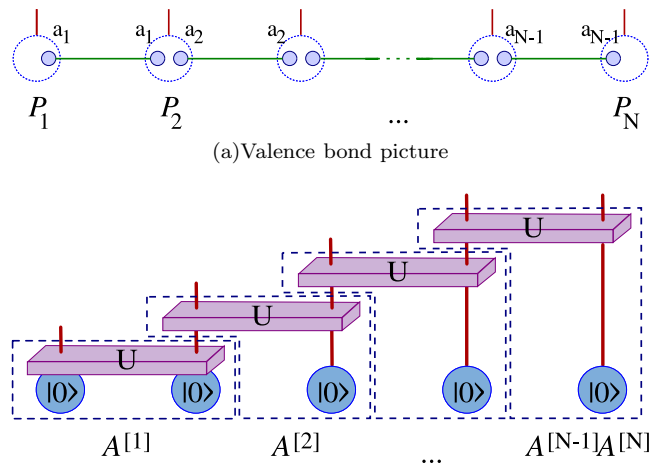

\subfigure[Valence bond picture]{
\label{fig:mps_seq-a}
\resizebox{\columnwidth}{!}{
\includegraphics{figures/mpsBasic.epsi}
}
}
\\
\subfigure[Sequential application of unitaries on
groups of $M+1=2$ sites]{
\label{fig:mps_seq-b}
\resizebox{\columnwidth}{!}{
\includegraphics{figures/mpsSeq.epsi}
}
}
\caption{Scheme of MPS construction. In~\ref{fig:mps_seq-a}, each entangled virtual pair is shown as a joined pair of circles, whereas the dashed circles represent the maps $P_k$ onto the physical spins, represented by vertical segments. In~\ref{fig:mps_seq-b}, each box represents the application of a unitary $U$ on two neighboring sites, and vertical lines correspond again to spin indices. The dashed lines show how $A$ matrices can be obtained from this sequential construction.}
\label{fig:mps_seq}
\end{figure}

Let us consider a chain of $N$ $d$-dimensional systems.
MPS are defined in several equivalent ways.
\bit
\item{\it{Valence Bond picture}.}
Each one of the physical spins is assigned two virtual particles
of dimension $D$, each of them sharing a maximally entangled state (bond),
$\sum_{a=1}^D | a, a\rangle$,
with their neighbor.
The state of the chain is obtained by applying at each site $k$ 
a map $P_k$
from the virtual pair onto the physical spin (see Fig.~\ref{fig:mps_seq-a}).
If the mapping on site $k$ is
$$
P_k=\sum_{i=1}^d\sum_{a,b=1}^D (A_k^i)_{a,b}|i\rangle\langle a, b|,
$$
the states constructed by this procedure have the form
\beq
|\Psi\rangle =\sum_{i_1,\ldots i_N=1}^d \mathrm{tr}(A_1^{i_1}\ldots A_N^{i_N})
|i_1,\ldots i_N \rangle,
\label{eq:mps}
\eeq
where each matrix $A_k^i$ has maximum dimension $D\times D$.

\item
{\it{Sequential generation}.}
As shown in~\cite{schon05seq}, an arbitrary MPS with bond dimension $D$ 
can be equivalently generated by the sequential application of unitary operations between 
an ancilla system of dimension $D$ and the physical 
sites of the chain. 
Alternatively, the use of the ancillary system can be substituted by the application
of unitary operations on sites of the chain, only, in a sequential manner
(see Fig.~\ref{fig:mps_seq-b}).
In this case, a unitary acting on $M+1$ sites can generate all MPS with bond 
dimension $d^{M}$.


\item
{\it{Effective site blocks}.}
Another possibility to construct a MPS is to
assign matrices $\tilde{A}_k^{\tilde{i}}$ to blocks of $M$ sites,
instead of individual sites.
The so constructed state is analogous to (\ref{eq:mps}),
\beq
|\Psi\rangle =\sum_{\tilde{i}_1,\ldots \tilde{i}_{N/M}=1}^{d^M} 
\mathrm{tr}(\tilde{A}_1^{\tilde{i}_1}\ldots \tilde{A}_{N/M}^{\tilde{i}_{N/M}})
|\tilde{i}_1,\ldots \tilde{i}_{N/M} \rangle,
\eeq
where now the sum runs over effective ``spin'' indices $\tilde{i}_k$ of 
dimension $d^M$.
This gives again a way to construct the
state, namely by applying unitary matrices, as described above, that
act sequentially on groups of adjacent blocks.

\eit

\subsubsection{Properties}
\label{subsec:mps_prop}

The most remarkable properties of MPS are the following.
\bit
\item{Basis for DMRG.}
MPS are intimately connected to DMRG and its success in the simulation
of large 1D quantum systems.
DMRG algorithms, in fact, 
optimize over MPS of fixed bond dimension, $D$, to 
approximate the physical state~\cite{schollwoeck05dmrg}.
In this sense, MPS provide a basis for a variational DMRG procedure.

\item{Efficiently contractable.}
The computation of expectation values of local operators in MPS can
be done efficiently.
Given an operator 
$O=O_1\otimes O_2\otimes\cdots\otimes O_N$, which is a tensor product of
local operators, its expectation value reduces to the trace of a matrix product 
\beq
\langle \Psi_{MPS}|O|\Psi_{MPS}\rangle = \mathrm{tr} (E_{O_1}^{[1]}\cdots E_{O_N}^{[N]}),
\eeq
where every term
$E_{O_k}^{[k]}=\sum_{i,i'} \langle i'|O_k|i \rangle [{A_k^{i'}}^* \otimes A_k^{i}]$
is a transfer matrix of size $D^2\times D^2$.

\item{Exponentially decaying correlations in the translationally invariant case.}
If we consider an infinitely long chain, 
described by a translationally invariant MPS, i.e.\ with the same $A$ tensor
for every site,
 generically the correlations between two sites decrease exponentially with 
the distance between them, 
$\langle O_k O_{k+\Delta} \rangle -
\langle O_k \rangle \langle O_{k+\Delta}\rangle \approxeq e^{-\Delta/\xi}$.
Here, $\xi$ is called the correlation length.

\item{Complete family.}
Any state of the Hilbert space for $N$ particles can be cast in the form of a MPS
with sufficiently large bond dimension ($D\approx {\cal O}(d^{N/2})$~\cite{vidal03eff,verstraete04dmrg}).
Thus, the MPS classify the whole state space according to the dimension $D$.
The lowest classes in this hierarchy prove to be most useful to 
describe the low energy sectors of quantum many-body systems with local interactions.

\item{Area Law.}
By construction MPS satisfy an area law, i.e.\ 
the entanglement entropy of a block of spins scales as the area of the block
boundary. 
In the case of a MPS with bond dimension $D$, as the boundary crosses only two bonds,
the entropy is upper bounded by $2 \mathrm{log} D$. 

\item{Parent Hamiltonian.}
Every MPS 
is the ground state of a local Hamiltonian.
Under some generic constraint on the MPS, this ground state is unique, and the
parent Hamiltonian is gapped~\cite{perez07mps}.

\item{Extensible to mixed states.}
The notion of MPS is extended from pure to mixed states
in the class of Matrix Product Density Operators (MPDO)~\cite{verstraete04mpdo},
which can be used to study one dimensional many-body systems at finite temperatures.

\eit

\subsection{Generalization to higher dimensions: PEPS}
\label{subsec:peps}

\subsubsection{Definition}
\label{subsec:peps_def}
\begin{figure}
\begin{tabular}{c}
\resizebox{\columnwidth}{!}{
\includegraphics{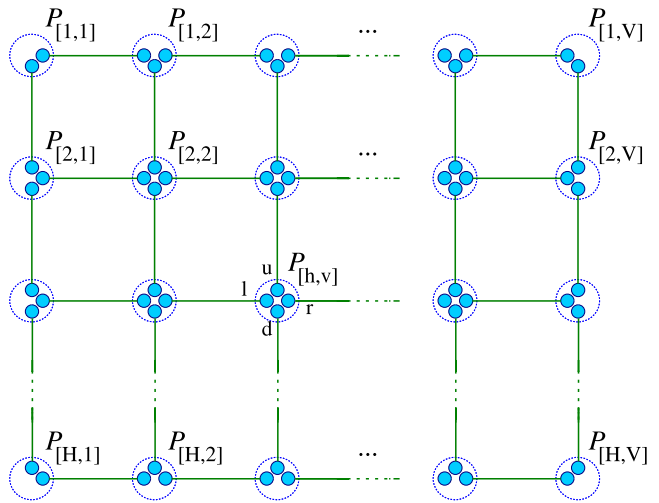}
}
\end{tabular}
\caption{Scheme of PEPS construction. Solid lines join pairs of virtual particles that
share a maximally entangled state, and dashed circles represent the mapping $P$ from the 
virtual particles onto the physical spin at each site.}
\label{fig:peps}
\end{figure}

The valence bond construction above can be generalized in a natural way
to graphs in higher dimensions by 
assigning to each site as many virtual particles as incoming edges.
This yields the definition of PEPS.
For example, a generic PEPS for a two dimensional $H\times V$ square lattice 
is constructed by representing 
each physical site $(r,\ c)$ by four auxiliary systems of dimension $D$, each of them
sharing a maximally entangled state with the adjacent neighbor, and 
then mapping all the virtual onto the physical degrees of freedom 
at each site (see Fig.~\ref{fig:peps}).
The state can be written 
$$
|\Psi\rangle=\sum_{i_{(1,1)}\ldots i_{(H,V)}=1}^d 
\mathrm{F}_2(\{ {B^{[r,c]}}^{i_{(h,v)}}\})|i_{(1,1)}\ldots i_{(H,V)}\rangle,
$$
where the four-index tensors $B^i$ contain the mapping between
virtual and physical systems at each site,
and the function $\mathrm{F}_2$ contracts all the virtual indices 
according to the bonds.

\subsubsection{Properties}
\label{subsec:peps_prop}

As a generalization of the MPS construction,
PEPS share with them some desirable characteristics.
In particular they are a complete set and satisfy an area law.
Nevertheless, PEPS can support large correlations
and cannot be efficiently contracted.
Next we detail these and other properties.

\bit
\item{Complete basis.}
As MPS, PEPS form a complete set, i.e.\ any state can 
be written as a PEPS with high enough bond dimension.

\item{No efficient contraction.}
Contrary to MPS, the cost of contracting PEPS scales in general exponentially with
the number of systems.
Therefore to devise a variational algorithm based on PEPS it is necessary to
use approximation methods~\cite{verstraete04peps,murg07hard}
or to restrict the 
variational set 
to a subfamily of states (as~\cite{bravyi08tensor} or the ones described below).

\item{Large correlations.}
Different to the case of MPS, the two-point correlation functions
in a PEPS do not have to decay exponentially with the distance between sites.
In~\cite{verstraete06crit}, it was shown that there exist PEPS
with very low bond dimension $D=d$ reproducing the correlations and 
expectation values of classical thermal states for any classical
two-body spin Hamiltonian. This is true in particular for
the classical Ising model at the critical temperature, which 
has algebraically decaying correlations.

\item{Area law.}
Like MPS, PEPS satisfy by construction the area law scaling of entanglement entropy,
since the maximum entropy of a block is determined by the number of broken bonds, i.e.\ 
the size of the boundary~\cite{verstraete04vbs}.

\item{Parent Hamiltonian.}
Each PEPS is the ground state of a local Hamiltonian.
If the PEPS satisfies an injectivity condition~\cite{perez07peps}, the ground
state is unique. Different to the case of MPS this does not suffice to
ensure the gapless character of the parent Hamiltonian.
\eit

\section{An alternative generalization: Sequential families}
\label{sec:upeps}

\subsection{Definition}
\label{subsec:upeps_def}

The central idea of the generalization proposed here is to extend 
the sequential construction scheme of MPS to 2D systems, 
by allowing also the application of unitary operations along a second 
dimension.
We can proceed in two ways.

\subsubsection{Sequentially Generated States (\SGS )}
\label{subsec:upeps1}
For a $H\times V$ lattice, we may define a new family of states in 
the following way.
We consider each row $r$ to be in a pure state described by a certain 
MPS of bond dimension $D$, defined by tensors $A^{[r,c]}$.
As already discussed, these tensors define a recipe for constructing
the row states by sequentially applying unitary operations on
groups of $M+1$ neighboring sites, with $M=\frac{\log D}{\log d}$.
Then we apply a second layer of unitary operations as follows.
Along each column, $c$, we apply unitary transformations on $M+1$ sites, 
starting on the $M+1$ bottommost rows and moving upwards, one row at a time 
(see Fig.~\ref{fig:upeps-a}).
Thus the first unitary applied on column $c$ is $U^{[H,c]}$,
whereas $U^{[r,c]}$ is the unitary operation acting on the physical index of
$A^{[r-M,c]}$ and on the uppermost $M$ spin indices after the application of
$U^{[r+1,c]}$.
As in the MPS case, the bond dimension $D$ along either the vertical or the
horizontal direction can be increased by 
applying unitary operations on a larger number of sites.

A well-known state that admits this description is the
cluster state~\cite{briegel01cluster}, which is given by the application of 
a single unitary to every pair of neighbours in the lattice.
In such case, applications of the unitary to different sites commute among themselves 
and thus we can apply the unitaries in the sequential order described above.

\subsubsection{Block Sequentially Generated States (\BSGS)}
\label{subsec:upeps2}

\begin{figure}
\subfigure[Construction of \SGS, with unitaries acting on $M+1=2$ rows]{
\label{fig:upeps-a}
\resizebox{\columnwidth}{!}{
\includegraphics{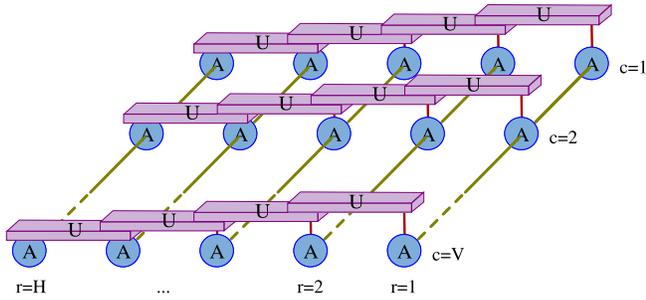}
}
}\\
\subfigure[Construction of \BSGS, with blocks of $N=2$ rows]{
\label{fig:upeps-b}
\resizebox{\columnwidth}{!}{
\includegraphics{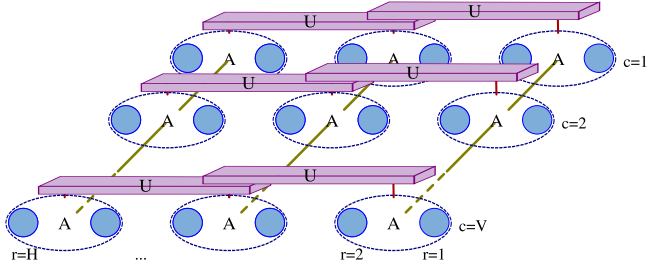}
}
}
\caption{Scheme of sequential construction. In~\ref{fig:upeps-a}, each tilted line represents 
a MPS, connecting $A$ tensors in a given row, whereas boxes $U$ correspond to the action of
a unitary on two neighboring rows. In~\ref{fig:upeps-b} the construction of a \BSGS ~is 
schematized.
The dashed ellipses represent effective blocks of sites, each of them described by a single 
tensor $A$. The tilted lines represent then MPS of larger dimension for a block of rows,
and the unitary boxes $U$ act now on groups of effective sites.}
\label{fig:upeps}
\end{figure}

Although, as described above, there is one natural way of extending the \SGS
~construction by using larger unitaries along either direction, 
it turns out that this procedure does not improve the descriptive power
of the \SGS ~family.
Instead, we
may think of another generalization, adapting the idea of MPS construction
from effective site blocks.
We must then define larger effective sites
on which the two layers of unitaries are then applied.
To this end, we first define a block taking together $N$ physical 
sites in the same column,
so that we are left with a $\frac{H}{N}\times V$ lattice where sites 
have physical dimension $d^N$.
On this system we construct a \SGS ~by first applying unitary operations
horizontally along each of the $\frac{H}{N}$ rows, to build MPS, and
then connecting the different rows by unitary operations on each column,
as described above. These unitaries now connect neighboring effective 
sites, i.e.\ blocks of spins (see Fig.~\ref{fig:upeps-b}).

Although the first definition~\ref{subsec:upeps1} is clearly contained in
the second one, the opposite is not true.
In fact, this second definition provides the systematic way to extend the 
family to cover the whole state space by taking larger blocks $N$, 
as we will discuss in the following paragraph.


\subsection{Properties}
\label{subsec:upeps_prop}

\bit
\item{Efficiently preparable and contractable.}
By construction, both families of sequentially generated states can be 
efficiently realized.
Their implementation requires only the sequential application of local 
unitary operations along horizontal and vertical directions.
 
To construct a \SGS ~state with bond dimension $D$ along both directions 
the unitary matrices must act on $M+1$ sites, with $D=d^M$.
For a $H\times V$ lattice 
the total number of unitary operations to
be applied is then $H (V-M)+ V (H-M)$.

Moreover, the contraction of \SGS ~can also be efficiently 
calculated on a classical computer.
It is easy to see that computing the norm of such a state reduces
to the product of norms of all the horizontal MPS, as the product of
all vertical unitaries appears contracted with exactly its
adjoint.
Therefore, the unitarity of vertical bonding matrices reduces the normalization
of the state to that of the underlying MPS.

The expectation value of the tensor product of a small number of local operators
can also be efficiently calculated.
Let us assume that we are interested in 
some tensor product of local operators acting on two sites, 
$(i_1,\,j_1)$, $(i_2,\,j_2)$,
$
\langle O \rangle=\langle O^{[i_1,j_1]}_1 \otimes O^{[i_2,j_2]}_2 \rangle.
$
The product of all the unitaries that act on a single column is itself a unitary
operation that commutes with all the others, and with local operators acting on different 
columns.
Therefore the contribution of all unitaries on columns different from $j_1$ and $j_2$
cancels in the expectation value.
The same is true for unitaries on columns $j_1$ and $j_2$ that only affect rows
above $i_1$ and $i_2$, respectively.
The expectation value can then be written as a product of norms of the rows above
times a contraction of a ladder structure. 
This is easily shown to scale as $d^2\,D^6$.~\footnote{The scaling 
depends exponentially on the number of local operators applied, so that
the calculation is feasible as long as this number is small.}

These arguments hold also for the \BSGS ~definition above,
with only the appropriate effective values of $d$, $H$ and $D$.

\begin{figure}
\resizebox{\columnwidth}{!}{
\includegraphics{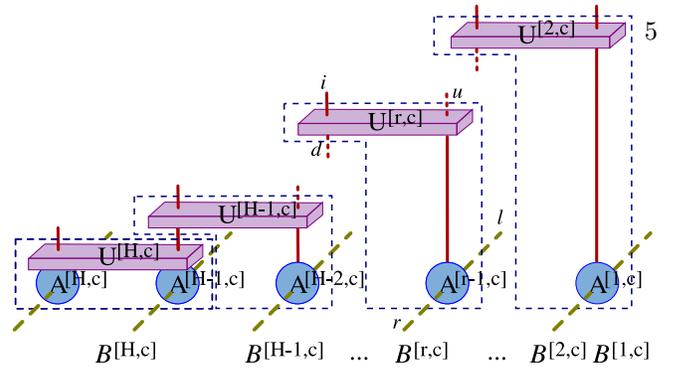}
}
\caption{Any \SGS ~can be written as a PEPS. The picture shows the construction 
of the $B$ tensors for a given column, $c$, in the particular case $M=1$
(i.e.\ unitaries acting on two rows). 
For a generic site, the tensor $B^{[r,c]}$ is thus determined by the MPS 
matrix $A$ corresponding to the site that lies $M$ rows above and the unitary 
that acts on this and the $M$ sites below.}
\label{fig:upepsB}
\end{figure}

\item{Subfamily of PEPS.}
Any \SGS ~state
can immediately be written as a PEPS of bond dimension upper bounded by $D$,
with tensors
${B^{[r,c]}}^i_{l u r d}=\sum_{j=1}^d {U^{[r,c]}}^{i u}_{d j}\, {A^{[r-M,c]}}^j_{l r}$.
Here $U^{[r,c]}$ is the unitary matrix that acts on rows $r-M$ to $r$ of column $c$,
and $A^{[r-M,c]}$ is the MPS tensor corresponding to row $r-M$.
The index $i$ is the free spin index of site $[r,c]$.
So, the PEPS gets the horizontal ($l$, $r$) bonds from the $r-M$ row,
while the vertical bonds $u$ and $v$ are the composition of $M$ spin indices 
corresponding to the upper or lower rows (see Fig.~\ref{fig:upepsB}).

The expression above is valid for rows $M<r<H$.
The $B$ tensors in the last row, $r=H$, would contain also the contraction 
with the physical indices of all the $A$ matrices corresponding to rows 
below $H-M$, 
${B^{[H,c]}}=U^{[H,c]}\, {A^{[H-M,c]}} 
\ldots {A^{[H,c]}}$.
On the other hand, the corresponding term for the
last unitary, ${U^{[M+1,c]}}^{i u}_{d j}\, {A^{[1,c]}}^j_{l r}$,
will contain the product of the $B$ tensors for rows $1$ to $M+1$, 
which can be obtained from this term by means of singular value 
decompositions.

The converse is not true, as an arbitrary \PEPS ~cannot always be 
expressed in this form.
To express an arbitrary $B$ tensor as a \SGS ~we could apply a singular 
value decomposition to split the horizontal indices, $lr$, from the rest.
This may in general yield up to $D^2$ singular values. If this number 
is larger than the physical dimension $d$, the result will not lead to
a valid $A$ matrix for a MPS with the same physical dimension.
This bound on the number of singular values of $B$ constitutes a 
necessary condition
for a state to be writable as a \SGS, but it is not sufficient. 
One needs also that the first part of the singular value decomposition admits a
reorganization of the indices to give a unitary matrix.

For the \BSGS, each product $U^{[r,c]}\, {A^{[r-M,c]}}$
will render a $B$ tensor corresponding to an effective block of sites.
To obtain a PEPS representation of the same state, such tensor
has to be decomposed as the contraction of the vertical indices
of $N$ individual $B$ tensors on the same column.
This can be always achieved by an adequate singular value decomposition.

\item{Decaying correlations.}
Due to their construction, which generalizes
the sequential generation of MPS, 
in a translationally invariant system these states show
correlations that decay exponentially with the distance along
both directions.
In this case, the translational invariance implies 
that the state is described by a single $A$ tensor and 
a single unitary operation.

Although the complete proof of this property can be found in the 
appendix, here we sketch the main ideas.
To check the property, we analyze independently correlations
along the horizontal and vertical directions.
In the first case,
we use the fact that, given the translational invariance, 
the underlying horizontal MPS states are 
exponentially close to a product, so that their two-point correlations 
within one row 
decrease exponentially with the distance. 
Only the second layer of unitaries 
acting along the vertical direction can introduce corrections
to this decay law. 
Nevertheless, we observe that such corrections
can increase only linearly with the number of rows in the system, 
so that the exponential decay dominates as the total size and the distance 
tend to infinity.

On the other hand, correlations between two sites of the same column
that lie on different rows
are only due to the second layer of unitaries, since in absence of the latter
the state is a tensor product of MPS states for each row.
In particular, for the kind of correlations under study,
the only contribution comes from unitaries acting on the single column involved. 
It is easy to see that
the situation is analogous to a translationally invariant MPS along the vertical direction
with larger effective site dimension.
This immediately implies that such correlations 
must also decay exponentially with the distance.



\item{Area law.}
Being a subfamily of PEPS, both \SGS~and \BSGS~satisfy the area law.

\item{Complete family.}
As discussed above, the family of \SGS ~cannot include arbitrary states, as in particular 
they are not always capable to describe a PEPS of 
given bond dimension.
Nevertheless, the family of \BSGS ~provides a way
of overcoming the limitations of the first one, by increasing the size of the 
effective blocks.
In this way, any state of a finite 2D system can be described as a \BSGS ~by
grouping together a large enough number of rows, $N$.
Notice that, in the limit, $N=H$ and the \BSGS ~description reduces
to a MPS describing a chain of $V$ $d^N$-dimensional sites. 

\item{Basis for a variational algorithm.}
The properties above make these families a suitable ansatz 
for a variational algorithm that looks for the ground state of local 
Hamiltonians.
The fact that they can be efficiently contracted grants the efficiency
of such procedure.
Although the first family cannot describe arbitrary states,
it is worth to explore its performance to find physically interesting 
states, arising as ground states of local Hamiltonians.
The second family, on the other hand, grants a systematic procedure
to improve the description of a system by considering larger and
larger blocks of sites.
The algorithm and the numerical study of its performance for both
sequential families are described in the following section.

\eit

\subsection{Variational algorithms}
\label{subsec:upeps_algo}

We consider Hamiltonians with  nearest neighbours interactions,
$H=\sum_{r,c}h_{[r,c]}$, where each term $h_{[r,c]}$ contains interactions
of site $(r,c)$ with its adjacent neighbours in both directions, as well as
possibly one-body terms.
The algorithm should minimize
the energy
$E=\frac{\langle \Psi | H |\Psi\rangle}{\langle \Psi |\Psi\rangle}$
with respect to the family of states in which we are interested.

The procedure can be built as in~\cite{verstraete04dmrg} 
by sequentially fixing all 
but one of the matrices ($A$ and $U$) that define the state, and then 
finding for  the free matrix the optimal
value, which minimizes the quantity above.
The cost of the algorithm will be determined by the cost of contracting 
the lattice for local operators, which, as already discussed, can be done 
efficiently.

We carry out this program in two phases. 
First, we apply the iterative procedure over all the $A$ matrices 
that define the
horizontal MPS by sweeping over each row from left to right and back.
This phase of the algorithm is almost identical to~\cite{verstraete04dmrg}.
As described there, a gauge condition that ensures normalization is 
applied at each step,
as well as techniques to improve the performance 
by storing partial contractions in memory as we move from one site to another. 
The only difference is that, in this case, different terms of the Hamiltonian 
contribute differently depending on the relative position of the rows on 
which they act with respect to the $A$ being optimized, and therefore
more terms need to be calculated and stored as going from one row to the
following.

The second phase of the algorithm consists in the optimization of the unitary
matrices by a similar procedure.
In this case, however, the quadratic character of 
$\langle \Psi | H |\Psi\rangle$
as a function of one particular $U$ is not enough to find the 
optimal matrices, 
because we must also impose the unitarity condition.
This cannot be done by applying a gauge transformation, as is done 
for the $A$ matrices, therefore we employ a slightly different approach
in order to find each $U$.
Instead of directly optimizing the quotient above, we apply a small variation
to the initial value of the unitary matrix, say
$U^{[r,c]}=e^{i\,\delta K} U^{[r,c]}_0$, where $\delta$ is a small 
real value and $K$ is an unknown  Hermitian matrix.
To the first order in $\delta$, the energy is a 
linear function in $K$ (and its adjoint)
that can be analytically optimized, with the constraints of hermiticity $K=K^{\dagger}$
and normalization $\mathrm{tr}(K^{\dagger}K)=1$.
For the so found value of $K$, we update the unitary 
(with the largest possible $\delta$)
and iterate the variation until convergence.

Contrary to the case of $A$ tensors, the latter perturbative procedure 
does not grant analytically  that 
a minimum is found at each step.
In practice, however, this approach
shows a good convergence.


\subsubsection{Numerical results}

To study the performance of this ansatz, we have implemented the 
algorithm above in Matlab, and applied it to the ground state
of different 2-body Hamiltonians, $H=\sum_{(ij)} h_{(ij)}$ on 2D lattices.

The tests have included random 2-body Hamiltonians,
where each $h_{(ij)}$ is a randomly chosen Hermitian operator acting on neighboring 
sites $(ij)$; 
the Heisenberg model,
$h_{(ij)}=\sigma_x^i \sigma_x^j +\sigma_y^i \sigma_y^j+\sigma_z^i \sigma_z^j$;
 and
a frustrated XX-model, 
$h_{(ij)}=J_{(ij)}\left(\sigma_x^i \sigma_x^j +\sigma_y^i \sigma_y^j\right)$,
with $J_{(ij)}=-1$ on every fourth edge (in both directions).
The algorithm has been run for
lattices up to size
$10\times10$,
and the results have been compared with those
obtained with PEPS~\cite{murg07priv}.

The following table shows the results for the \SGS
~family.
For the models mentioned above and the specified lattice sizes, 
the table contains the lowest energy found with this algorithm, $E_0$,
using bond dimension $D$~\footnote{MPS have bond dimension $D$, 
and unitaries act on $M+1$ sites, so that $d^M=D$.},
together with the relative error, $\varepsilon_r$, with respect to the PEPS result
(obtained with 
 with bond dimension $D=4$ and time step $\delta t=0.001$).

\begin{center}
\begin{tabular}{|c|c|c|c|c|}
\hline
model & lattice & $D$ & $E_0$ & $\varepsilon_r$ \\
\hline
\multirow{3}{*}{random} & \multirow{3}{*}{8$\times$8} & 
2 & -169.309 & 5.7$\times 10^{-3}$\\
\cline{3-5}
 &  & 4 & -169.556 & 4.3$\times 10^{-3}$\\
\cline{3-5}
 & & 8 & -169.613 & 4.0$\times 10^{-3}$\\
\hline
\multirow{6}{*}{Heisenberg} & \multirow{3}{*}{8$\times$8} & 2 & -153.737 & 0.0254\\
\cline{3-5}
 &  & 4 & -154.031 & 0.0235\\
\cline{3-5}
 &  & 8 & -154.142 & 0.0228\\
\cline{2-5}
& \multirow{3}{*}{10$\times$10} & 2 & -244.830 & 0.0209\\
\cline{3-5}
 &  & 4 & -245.244 & 0.0193\\
\cline{3-5}
 &  & 8 & -245.383 & 0.0187\\
\hline
\multirow{3}{*}{frustrated XX} & \multirow{3}{*}{8$\times$8} & 2 & -90.598 & 0.016\\
\cline{3-5}
 & & 4 & -91.242 & 9.0$\times 10^{-3}$\\
\cline{3-5}
 & & 8 & -91.398 & 7.3$\times 10^{-3}$\\
\hline 
\end{tabular}
\end{center}

We observe that in all cases a good precision is attained
with very low bond dimension, and increasing $D$ does not 
significantly improve the result, contrary
to what occurs with MPS or general \PEPS.
This can be readily understood from already discussed arguments.
Any \SGS ~with a fixed bond dimension can be 
represented as a PEPS of the same virtual dimension, but not the other 
way round.
The rank of the singular value decomposition of the PEPS tensor
should be smaller than $d$ for it to yield a valid MPS of the same 
physical dimension.
And this restriction is independent of the bond dimension allowed for the \SGS. 
Hence the ground state of these Hamiltonians can
only be approximated to a finite precision with a \SGS ~of physical 
dimension $d$.

However, constructing the sequential state from blocks instead of individual sites
allows us to get closer to the true ground state  
by considering larger effective sites,
and therefore larger effective $d$.
We have applied the same algorithm using this \BSGS ~family as
ansatz, with effective sites of $N=2$ rows.
As the following table shows, this reduces the relative error in more than 30\%
in all the cases under study ($D$ indicates now the bond dimension 
for the MPS, which was chosen equal to the effective dimension, $d^2$).

\begin{center}
\begin{tabular}{|c|c|c|c|c|c|}
\hline 
model & original lattice &$N$& $D$ & $E_0$ & $\varepsilon_r$ \\
\hline
random & 8$\times$8 & 2 & 
4 & -169.963 & 1.96$\times 10^{-3}$\\ 
\hline 
\multirow{2}{*}{Heisenberg} & 8$\times$8 & 2 & 
4 & -155.231 & 0.0159\\ 
\cline{2-6}
& 10$\times$10 & 2 & 
4 & -246.852 & 0.0128 \\ 
\hline
frustrated XX & 8$\times$8 & 2 & 4 & -91.703 & 4.03$\times 10^{-3}$\\ 
\hline
\end{tabular}
\end{center}

\section{Discussion}
\label{sec:conclu}

The study of efficient representations of quantum many-body states 
has been very successful in the description of one dimensional systems.
Quantum information has provided a different perspective to understand
these 
techniques~\cite{latorre03crit,vidal03eff,schollwoeck05dmrg,hastings06gap,hastings07entropy,eisert06dmrg}.
The hierarchy of MPS yields an adequate variational class of states
for numerical methods in one dimension.
In higher dimensions there are theoretical and practical limitations
to the application of these states or 
their natural generalization.
Therefore it is interesting to look for alternative representations that
are capable to describe the low energy sector of physically relevant systems
while offering better contractability.

Here we have presented extensions of the sequential construction of MPS
to two dimensions which can be introduced using two different approaches.
Both families defined here represent subsets of states which can be efficiently 
prepared in practice,
as their definition immediately provides a sequential recipe 
for its preparation.
We have proved that those states show exponentially decaying correlations.
Moreover, their contraction can also be done efficiently, which makes them 
a good ansatz for variational procedures.
We have numerically studied the performance of both approaches as
ground states for different local Hamiltonians.
The results show that, although the suitability of the 
first approach is limited, the second one provides a systematic way of
approximating the ground state of these systems.
Our tests have only shown the feasibility of this second approach,
which nevertheless has higher requirements from the computational
point of view.

As already discussed, the different performance of both families is due 
to the fact that,
for a PEPS to be writable as a \SGS ~of dimension $d$,
it is necessary (although not sufficient) that the 
singular value decomposition splitting the $lr$ PEPS horizontal indices
from the rest has at most $d$ values different from zero.
From a different point of view, it is sometimes possible to write a given 
\PEPS ~with more than $d$ singular values in the decomposition above,
as a certain local projection of a \SGS ~state of larger physical dimension
onto a $d$-dimensional subspace for each site.
It is the case of the toric code~\cite{kitaev03toric}, which has a 
\PEPS ~description of $D=2$~\cite{verstraete06crit}.
The $B$ tensor has four non-zero singular values according to the
decomposition $iud$ vs $lr$. If we take the right part of this decomposition 
as a MPS description for a chain of dimension $d'=4$, it turns out to be 
possible to complete the rest of the decomposition onto a $8\times 8$ 
unitary matrix.
This would correspond to free indices of dimension $d'$
from which we could recover the physical $d=2$ by projecting onto a
local subspace.
 
Note that even if the rank of the singular value decomposition of $B$
is smaller than $d$, it is not always possible to find a \SGS ~description
for the state. 
For example, the PEPS constructed in~\cite{verstraete06crit} 
that reproduces the correlations and 
expectation values of thermal states for the classical 
Ising model has very low bond dimension $D=d$. From its explicit 
representation it is possible to see that the 
rank of the relevant singular value decomposition is only $d=2$. 
However, it is not possible in general to 
represent this state as a \SGS.

\appendix*
\section{Proof of the decay of correlations}
\label{sec:app}

We want to calculate a correlation function of
the form
$\langle O^{[h_1,v_1]}_1 O^{[h_2,v_2]}_2\rangle -
\langle O^{[h_1,v_1]}_1\rangle \langle O^{[h_2,v_2]}_2\rangle
$ on an infinitely large 2D lattice with translational symmetry.

The decay of correlations along both directions
follows from the same property of MPS. 
Let us first calculate the decay in the vertical direction, i.e.\ that of
application of the unitaries.
Notice that if $v_1=v_2=v$, each of the expectation values above can be written
as
$$
\mathrm{tr}(O\,U^{[h]}\ldots U^{[H]} \rho^{\otimes H}_{MPS} {U^{[H]}}^{\dagger}\ldots {U^{[h]}}^{\dagger}),
$$
where $\rho_{MPS}$ is the reduced density matrix for the single site 
occupying column $v$ in any of the rows, and $h$ is the first row
touched by the corresponding operator $O$. 
Since the state of every
row is described by the same MPS, all such reduced density matrices are
the same.
Therefore, we are left with a tensor product of $H$ identical 
single site density matrices
connected by the sequence of unitaries that act on column $v$.
Such a construction, where all the degrees of freedom but the ones on 
column $v$ have been traced out, can be written as a 
MPDO~\cite{verstraete04mpdo}
$$
\rho=\sum_{i_k,\ i_k'}\mathrm{tr}({M^{[h_1]}}^{i_1i_1'}\ldots{M^{[H]}}^{i_Hi_H'})
|i_1\ldots i_H\rangle \langle i_1'\ldots i_H'|,
$$
 with matrices of the form
\footnote{For the bottommost and uppermost sites, the matrices are 
built in slightly different form, to respectively include the lowest MPS and split
the $M$ uppermost physical indices.
These considerations do not affect the following argument.} 
$${M^{[r]}}^{i\, i'}_{(\alpha \alpha'),(\beta \beta')}=
\sum_{\gamma\ \gamma'} {U^{[r]}}_{\alpha\gamma}^{i\beta}\rho_{\gamma \gamma'} {{U^{[r]}}^{*}}_{\alpha'\gamma'}^{i'\beta'}.$$
The correlation function for operators $O_1$ and $O_2$ acting on rows 
$h_1$ and $h_2$ can then be calculated using the purification of this MPDO, 
as a (translationally 
invariant) MPS of physical dimension $d^2$, for which the correlations decay
exponentially with the distance $h_2-h_1$.


If the operators are placed along the horizontal direction, $h_1=h_2=h$,
the expectation value $\langle O^{[h,v_1]}_1 O^{[h,v_2]}_2\rangle$
takes a similar form
\beq
\langle O\rangle=\mathrm{tr}(O_1\otimes O_2 \,\tilde{U}^{[h_1]}\ldots \tilde{U}^{[H]} \tilde{\rho}^{\otimes (H-h_1+1)}_{MPS} (\tilde{U}^{[H]})^{\dagger}\ldots (\tilde{U}^{[h_1]})^{\dagger}),
\label{eq:exphor}
\eeq
where each reduced density matrix, $\tilde{\rho}^{[r]}_{MPS}$,
corresponds now to two physical sites of row $r$, those
on columns $v_1$ and $v_2$.
As the two-sites reduced density matrix of a translationally 
invariant MPS, it is exponentially
close to a product of single-site density matrices.
This can be seen by writing its explicit form,
$$
\tilde{\rho}^{[r]}=\mathrm{tr}(E_{\bf 1}^{[r,1]}E_{\bf 1}^{(v_1-2)}
A \otimes A^{\dagger}
E_{\bf 1}^{(v_2-v_1-1)}A \otimes A^{\dagger} E_{\bf 1}^{(V-v_2-1)}
E_{\bf 1}^{[r,V]}
).
$$
Here $E_{\bf 1}$ are the $D^2\times D^2$ transfer matrices which, due to the 
translational invariance, do not depend on the column, except for the 
first and last sites of each row.
Under some generic condition on matrix $E_{\bf 1}$~\cite{perez07mps}, the
$v_2-v_1-1$ power of this matrix can be approximated by a product when the distance
$v_2-v_1$ becomes very large, so that
\beq
\tilde{\rho}\approxeq \sigma^{[v_1]}\otimes \sigma^{[v_2]} +
\varepsilon 
\tilde{\sigma}^{[v_1v_2]},
\eeq
where $\sigma^{[k]}$, $\tilde{\sigma}^{[k]}$ act on a single system each, 
and the factor $\varepsilon=\left(\frac{\lambda_2}{\lambda_1}\right)^{v_2-v_1-1}$ is 
determined by the ratio of the second largest 
eigenvalue of $E_{\bf 1}$ to the largest one, and decays exponentially fast with 
the distance $v_2-v_1$.
Moreover, if the edges of the lattice are infinitely far away, i.e.\  
$v_1,\, V-v_2\rightarrow \infty$, then $\sigma^{[v_1]}=\sigma^{[v_2]}$ (with a 
global normalization factor).


Since each row density matrix is exponentially close to a product,
\beq
\| \tilde{\rho} - \sigma^{[v_1]} \otimes  \sigma^{[v_2]}\|_1 \leq \cal{O}(\varepsilon),
\eeq
one can show by induction that this is also the case for the $H'$-fold 
tensor product appearing in~\ref{eq:exphor} ($H'=H-h_1+1$),
\beq
\| \tilde{\rho}^{\otimes H'} - (\sigma^{[v_1]})^{\otimes H'} \otimes 
(\sigma^{[v_2]})^{\otimes H'} \|_1 \leq {\cal O}(H' \varepsilon).
\eeq

The unitary matrices $\tilde{U}$ appearing in (\ref{eq:exphor})
are the tensor products
of a unitary matrix acting on the corresponding row and column $v_1$ 
times the one on column $v_2$,
$
\tilde{U}^{[r]}=U^{[r,\ v_1]}\otimes U^{[r,\ v_2]}
$.
Acting with them on the tensor product above does not increase the
trace norm, and moreover respects the tensor product structure of the second term,
so that the reduced density matrix corresponding to columns $v_1$ and $v_2$
satisfies
\beq
\| \tilde{\tau}^{[v_1 v_2]} - \tau^{[v_1]} \otimes  \tau^{[v_2]}\|_1 
\leq  {\cal O}(H' \varepsilon).
\eeq
Therefore it is easy to show that, for any pair of operators $O_1$, $O_2$
each one acting on $\mathds{C}^d$, and with operator norm bounded by 1,
\bea
|
\mathrm{tr}(O_1\otimes O_2 \tilde{\tau}^{[v_1 v_2]}) &-& \mathrm{tr}(O_1\otimes{\bf 1} \tilde{\tau}^{[v_1 v_2]})
\mathrm{tr} ({\bf 1} \otimes O_2  \tilde{\tau}^{[v_1 v_2]})
|
\nonumber
\\
&\leq&
4 {\cal O}(H' \varepsilon).
\eea
Thus the correlations decrease exponentially with distance 
\footnote{In other words, in order to get a contribution from the multiple rows, $H$, 
that compensates the 
exponential decay of MPS correlations, the size of the system should increase 
exponentially fast.
But given a fixed large size, the correlations will suffer an exponential decay.}.


Under some additional condition, it is possible to obtain a tighter bound on the correlations.
To this end we can define again for a row $r$ the matrices
$${M^{[r,v]}}^{i\,i'}_{(\alpha \alpha'),(\beta \beta')}=
\sum_{\gamma\ \gamma'} {U^{[r,v]}}_{\alpha\gamma}^{i\beta}\sigma^{[v]}_{\gamma \gamma'} {{U^{[r,v]}}^{*}}_{\alpha'\gamma'}^{i'\beta'},$$
for $v=v_1$, $v_2$, 
and an analogous $\tilde{M}$ matrix using $\tilde{\sigma}$ and $\tilde{U}^{[r]}$,
so that the reduced density matrix for sites $[h,v_1]$, $[h,v_2]$ can be written as
\beq
\rho_{\tilde{i}\tilde{j}}=\mathrm{tr}(M_{[h]}^{\tilde{i}\tilde{j}}\cdot 
G^{(H-h-1)}\cdot G_{[H]}),
\label{eq:redrhoH}
\eeq
where
\bea
M_{[h]}&=& U^{[h,\ v_1]}\otimes U^{[h,\ v_2]} \tilde{\rho}
(U^{[h,\ v_1]})^{\dagger}\otimes (U^{[h,\ v_2]})^{\dagger}
\nn
\\
&\approxeq&
M^{[h,v_1]}\otimes M^{[h,v_2]}+\varepsilon
\tilde{M}^{[h,v_1 v_2]},
\eea
and $\tilde{i}$ ($\tilde{j}$) are the double indices $i\,i'$ ($j\,j'$) appearing in 
the $M$ matrices for each site.
These indices are traced out on rows where no operator acts, yielding
\bea
G&=&\sum_{\tilde{i},\,\tilde{j}}\delta_{i\,i'}\delta_{j\,j'}M_{[h]}^{\tilde{i}\tilde{j}}
\nn
\\
&\approxeq&
G^{[v_1]}\otimes G^{[v_2]}+\varepsilon
\tilde{G}^{[v_1 v_2]},
\eea
where each individual $G^{[v]}$ is obtained by tracing out the physical index in $M^{[h,v]}$.

If the matrices $G^{[v_1]}$ and $G^{[v_2]}$ have a single maximal eigenvalue, with 
multiplicity one, also the whole matrix $G$
will have a single maximal eigenvalue $\tilde{\mu}$, with the same multiplicity, so that
$$G^{(H-h-1)}\approxeq (\tilde{\mu})^{(H-h-1)} |\tilde{\mu}^L\rangle\langle\tilde{\mu}^R |,
$$
plus terms that decrease exponentially with $H$.
The leading eigenvectors, using perturbation theory~\cite{kato95}, will be given by 
the product of eigenvectors of $G^{[v_1]}$ and $G^{[v_2]}$, plus some contributions
of order $\varepsilon$ whose number is bounded by the dimension of the (finite) matrices $G$.

On the other hand, the last matrix in (\ref{eq:redrhoH}), $G_{[H]}$, is the corresponding $G$ matrix for 
the bottommost unitary, which
therefore includes, instead of a single row, the tensor product of all the last $M+1$
rows, and then contributes at most with $\varepsilon (M+1)$ terms to the corrections.
\bea
G_{[H]}&=&U^{[H,\ v_1]}\otimes U^{[H,\ v_2]} (\tilde{\rho})^{\otimes M+1}
(U^{[H,\ v_1]})^{\dagger}\otimes (U^{[H,\ v_2]})^{\dagger}
\nn
\\
&\approxeq &
G_{[H]}^{[v_1]}\otimes G_{[H]}^{[v_2]}
+\varepsilon \sum_{k=0}^{M}(\sigma^{[v_1]})^k \tilde{\sigma}^{[v_1]}(\sigma^{[v_1]})^{M-k}
\nn
\\
&&
\otimes
(\sigma^{[v_2]})^k \tilde{\sigma}^{[v_2]}(\sigma^{[v_2]})^{M-k}.
\eea

Therefore, under this no-degeneracy assumption for matrices $G$ (which seems to be generic, 
after some numerics) the corrections to the tensor product 
structure of (\ref{eq:redrhoH}) are of order
$\varepsilon (1+{\cal O}(D^4)+{\cal O}(M))$, which does not depend on the size of the system.



\acknowledgments
We thank V. Murg for providing the numerical results from PEPS algorithms.
This work was supported by EU Strep Compas,
by DFG through Excellence Cluster MAP 
and FOR 635,
and by Spanish grant MTM2005-00082.

\bibliography{sgpeps}

\end{document}